\documentclass[aps, prb, reprint, superscriptaddress]{revtex4-2} 

\usepackage{graphicx, amsmath, amssymb}
\setcitestyle{super}

\begin{document}

\title{Letter to the Editor:  Diet Soda and Liquid Nitrogen}

\maketitle

In a Letter to the Editor\cite{liljeholm:2009} regarding an article by Coffey\cite{coffey:2008}, Anders Liljehom pointed out that iron filings are an attractive substitute for Mentos in the popular diet soda and Mentos reaction.  We would like to share our discovery of an additional method to produce the reaction: the direct immersion of an open plastic soda bottle into liquid nitrogen.  In this case, the strong cooling through the thin container wall leads to rapid nucleation.  

Because no material is introduced into the liquid, this method has the pedagogical advantage of illustrating that the diet soda and Mentos reaction comes from the rapid release of carbonation alone.  This method also has the practical advantage that the nozzle does not need to be designed to allow the introduction of a catalyst into the liquid.  The bottle only has to be partially immersed to produce the reaction.  Dry ice or another cryogen might work as a substitute for liquid nitrogen.

Liquid nitrogen and other cryogens are frequently used in scientific demonstrations.  This discovery raises an additional safety warning about working with cryogens, which is that carbonated liquids can quickly produce dangerous pressures in containers that come into contact with cryogens.  This method should not be used with glass bottles or closed containers.  \\

\hfill Bart H. McGuyer, 

\hfill Justin M. Brown, and 

\hfill Hoan B. Dang 

\hfill \textit{Graduate Students}

\hfill \textit{Department of Physics}

\hfill \textit{Princeton University}

\hfill \textit{Princeton, New Jersey 08544}

\end{document}